# Universality in the Magnetic Response of Metamagnetic Metals

B.S. Shivaram[1]*, D.G. Hinks[2], M. B. Maple[3], M.A.deAndrade[3] and P. Kumar[4]

[1] Department of Physics, University of Virginia, Charlottesville, VA. 22901.
[2] Materials Science and Technology, Argonne National Labs, Argonne IL. 60637
[3] Department of Physics, University of California, San Diego, La Jolla CA. 92093 and
[4] Department of Physics, PO Box 118440, University of Florida, Gainesville, FL. 32611

(17 February, 2014)

### ABSTRACT

We report in this paper, measurements of the nonlinear susceptibility $\chi_3$ (T) in the metamagnetic heavy fermion (HF) compound $UPt_3$. At high temperatures, $\chi_3$ (T) < 0 and small. It turns positive for T ≤ 35K, forms a peak at T ≅ 10K and then decreases to zero with further decreasing temperature. The peak in $\chi_3$ occurs at a temperature $T_3$ roughly half of $T_1$, the temperature of the maximum in the linear susceptibility. We present results on $URu_2Si_2$ and $UPd_2Al_3$ to show that this feature is common to other HF materials. A two level model to describe the metamagnetic transition, with separation between the levels being the only energy scale, captures all experimentally observed features.

*bss2d@virginia.edu - to whom all correspondence should be addressed.

Given the myriad ways in which one can arrange atoms to form crystal structures and the possibilities of placing electrons in them the discovery of a common behavioral pattern in the electronic properties often leads to new microscopic insights. In metals, recent discoveries of universal rules such as the Kadowaki-Woods relation[1] which connects the electron effective mass enhancement to the temperature dependence of the resistivity in Fermi liquid systems, the constancy of the Wilson ratio[2], the Homes plot[3] and other similar scaling laws have assisted in identifying the dominant energy scales, and helped in insights to build microscopic theories.

Our purpose in this paper is to show that the magnetic response of many electronic materials that undergo a metamagnetic transition also exhibit universal features in their nonlinear susceptibility. Along with a jump in the magnetization at a critical field, $H_m$, metamagnetism also entails a positive nonlinear susceptibility at low temperatures (it is negative at high temperatures), a peak in the linear susceptibility $\chi_1(T)$ at a temperature $T_1$, a peak in the leading order nonlinear susceptibility $\chi_3(T) > 0$ at a lower temperature $T_3$. These are features widely seen in different materials with diverse lattice structures and electronic properties. Hirose et. al[4] have noted the correlation between $H_m$ and the temperature $T_1$ of the maximum in $\chi_1(T)$. Goto et. al pointed out earlier that the metamagnetic field $H_m$ also correlates with the value of $\chi_1(T)$ at its maximum[5]. Both of these correlations, found in a number of materials, indicate a dominant presence of one energy scale. In the following we report a new correlation, one between the peak temperatures $T_3$ for $\chi_3(T)$ and $T_1$ namely $T_3 = T_1/2$. We also present a new model that captures all of these correlations.

For purposes of illustrating this universality we consider the heavy electron class of materials. Within these systems metamagnetic behavior is seen in several uranium compounds[6,7,8,9,10], cerium compounds[11,12,13,14], as well as in the ytterbium compounds[15]. Most of these systems possess either hexagonal or tetragonal symmetry. The same behavior is also observed in the skutterudite structure in $PrOs_4Sb_{12}$ a moderately heavy electron system[16] which is close to being cubic[17].

To analyze the temperature dependence of the magnetic response in these metals it is customary to perform a Curie Weiss fit to the measured linear susceptibility in the 'high temperature' limit and extract an effective magnetic moment. A peak in the linear susceptibility $\chi_1$ universally found in these materials is a deviation from such a fit and is often understood in terms of crystalline electric fields. Irrespective of whether these systems order magnetically (e.g. into an antiferromagnetic state) at lower temperatures there are crystal field determined sequence of energy levels which determine the thermodynamic observables. Heat capacity data is often used to narrow down on a specific crystalline electric field (CEF) level scheme which is then used to evaluate the magnetic susceptibility[18] $\chi_1$(or vice versa[19]). This approach places emphasis on system specific CEF levels. However many universal physical effects are at play in the d and f electron based itinerant electron systems.  In such systems presence of partial Kondo screening of magnetic moments in a lattice and exchange between moments on different sites imply coexistence of both long range order and local single ion effects.  We are able to show in the following that a model with a single energy scale is capable of incorporating these physical effects and in describing all of the above noted correlations in the magnetic properties in a comprehensive manner.

To analyze the field dependence of the equilibrium magnetization, M we can write:

$$M(T) = \chi_1(T)B + \chi_3(T)B^3 + ....$$
(1)

In practice the parameter $\chi_3$ is extracted from the slope of the straight line in a plot of M/B vs. $B^2$.  Since M must eventually saturate, $\chi_3 < 0$, unless there is an instability towards a larger magnetization as in metamagnetism.  A small parasitic constant magnetization, $M_0$, invariably present in all experiments is subtracted to extend the linear region in such a plot.  In the heavy electron class of materials since the magnetization is enhanced by nearly two orders of magnitude compared with 'ordinary' paramagnets the experimental determination of $\chi_3$ is especially convenient.  In this class of materials measurements of $\chi_3$ for $URu_2Si_2$, $CeRu_2Si_2$, $UPd_2Al_3$, $PrOs_4Sb_{12}$, $UBe_{13}$, $U_2Zn_{17}$ and $(U,Th)Be_{13}$ have appeared so far[20,21,22,23,24].  In the case of $URu_2Si_2$ the focus has been on understanding the novelty of the physics in the vicinity of the hidden order transition which occurs at $T_{HO}$=17 K.  $\chi_3$ in this case exhibits a very sharp peak at 17.5 K.  In the other systems where there is a definite magnetic order, $U_2Zn_{17}$ and $(U_{1-x},Th_x)Be_{13}$, an increase in $\chi_3$ to large  positive values at the magnetic transition is observed.  For the purposes of this paper the focus is rather on the behavior of both the linear and nonlinear susceptibility over a broad temperature region and thus the sharp behavior near magnetic transitions is ignored.  Rather, our goal is to understand the new measurements on $UPt_3$ presented below as well as to present a context in which existing measurements in a variety of systems can be understood.

In the main part of figure 1 we show the measured values of $\chi_3$ for $UPt_3$ for the field applied along the a-axis of the hexagonal crystal.  The data shown were obtained by measuring magnetization M in fields to 5 T in a Quantum Design MPMS SQUID based magnetometer on high quality single crystals grown by vertical float zone refining.  The values of $\chi_1$ and $\chi_3$ are obtained as the intercept and the slope of M/B vs. $B^2$ plots respectively at different temperatures.  Our values of $\chi_1$, which exhibits a peak at 20 K, are in excellent agreement with those published previously.  There is also a peak in $\chi_3$ centered at 10 K which is half the temperature where the peak in $\chi_1$ occurs.  The earlier reports in $PrOs_4Sb_{12}$ and $CeRu_2Si_2$ where $T_3$ and $T_1$ have been measured are consistent with the ratio of $T_3/T_1$ in $UPt_3$. (Refs. 24and 22).  In $CeRu_2Si_2$, doping the Ce site with Yttrium shifts $T_1$ substantially and $T_3$ follows suit precisely at half the value. Of the other systems where measurements in $\chi_3$ are available, $UBe_{13}$, a cubic system is particularly notable - it has no peak in $\chi_1$ or $\chi_3$. We summarize all these observations in figure 2, where we plot $T_3$ vs. $T_1$. The linear correlation between the two temperatures is apparent in this figure. In such a plot $UBe_{13}$ belongs at the origin[25]. In the measurements presented above the maximum applied field (5T) is significantly less than the field where metamagnetism sets in.

Metamagnetism, was originally proposed by Wohlfarth[26] for an intinerant electron system as a Fermi surface instability. More recently, there have been discussions by Vollhardt for a Hubbard model (in the context of a Gutzwiller solution)[27],[28] and by Bedell[29] and others in the context of Fermi liquid theory and the magnetic field dependence of Landau's Fermi liquid parameters. More recent work in a microscopic description of metamagnetism in heavy fermions can be seen in Kusminsky et. al.[30] and by Spalek and coworkers[31]. Using more modern tools such as dynamical mean field theory with numerical renormalization group and within the context of a half filled Hubbard model Bauer has also studied itinerant metamagnetism[32].

In a metamagnet clearly the ground state moment must be small. There should however be an excited state, with a large moment that moves down in energy with increasing magnetic field. Thus a two level system should be sufficient to account for all of the observed properties with the field induced level crossing corresponding to metamagnetism. Consider a two level system, separated by an energy $\Delta$, where the effect of magnetic field is to reduce $\Delta$. The lower of the two levels has a smaller magnetization than the upper one. It is well established that the ground state for the single impurity problem is a singlet and therefore it is reasonable to assume that this is the case even for the Kondo lattice. A single site energy scale Hamiltonian in the presence of a magnetic field B, for a pseudo spin S = 1, could be written as,

$$H = \Delta S_z^2 - \gamma S_z B \qquad\qquad (2)$$

At B = 0, the lower energy level corresponds to $S_z = 0$ and the upper one corresponds to $S_z = \pm 1$. Here $\gamma$ is the putative gyromagnetic ratio containing information, inter alia, about the microscopic details such as the J value, the size of the moment etc.. The model can be seen to lead to a partition function Z = 1+2exp (-1/$\tau$) cosh ($\gamma$B/$\Delta\tau$). Here $\tau$ = kT/$\Delta$ is the temperature scaled to $\Delta$. The magnetization is given by M/$\gamma$ = 2sinh b/$\tau$ /[$e^{1/\tau}$ + 2cosh b/$\tau$] with b = $\gamma$B/$\Delta$. The magnetization rises with a maximum derivative with respect to the magnetic field at b = 1. When expanded in powers of b, the susceptibilities are given as,

$$\chi_1 = \frac{\gamma^2}{\Delta} \frac{1}{\tau} \frac{1}{1+\frac{1}{2}e^{\frac{1}{\tau}}} \qquad\qquad (3)$$

$$\chi_3 = \frac{\gamma^3}{\Delta^3} \frac{1}{6\tau^3} \frac{\frac{1}{2} e^{1/\tau} - 2}{(1 + \frac{1}{2} e^{\frac{1}{\tau}})^2} \qquad (4)$$

These two functions are shown in the top panel of figure 3. The linear susceptibility has a maximum at $\tau_1 = 0.69$ while the nonlinear susceptibility has a maximum at $\tau_3 = 0.27$. The ratio of these two temperatures, $\tau_3/\tau_1 = 0.4$. The observed peak temperatures ratio of 0.5 and the experimentally observed scaling of the metamagnetic field**Error! Bookmark not defined.** support the simple model here. The zero field and low temperature ground state is nonmagnetic. The excited state energy level splits due to the magnetic field. One branch comes down. The $S_z = -1$ branch intersects the $S_z = 0$ branch at the "metamagnetic" field and takes over as the ground state with a larger magnetization. As suggested by the minimal model proposed above there is only one energy scale that governs all temperature dependences.

In this model, (as in most experiments) there is no "transition" at the metamagnetic field rather there is a rapid rise in the magnetization, which is also related to a positive $\chi_3$. A phase transition can be brought about by introducing a self consistent field, such as one that arises due to an exchange interaction. The resulting features will modify the details of the observables. The materials discussed here though seem to belong to a class where the onsite interaction as provided by the crystal electric fields are nominally sufficient to understand all correlations based on a single energy scale, namely the separation between the onsite energy levels. However, interaction effects do play a role. It is well known that the d and f electrons are hybridized with the s electrons and play a dominant role in the thermodynamic properties of metamagnets. Their interaction shows up in a variety of ways, most prominently in the form of a reorganization of the onsite energy levels. One detailed analysis (albeit in the framework of a molecular field approximation) of various effects was carried out by Morin and Schmitt[20] (MS). MS started with the correct (depending on the crystal structure) crystal field energy levels and proceeded to highlight several critical benchmark effects which arose as they included the exchange interaction and quadrupolar interaction effects. Bauer et al[24] calculated the nonlinear susceptibility of $PrOs_4Sb_{12}$ (in their study of quadrupolar fluctuations) starting with the correct crystal and electronic configurations. They found a peak in both $\chi_1$ (T) and $\chi_3$ (T) with magnetic fields in the 001 and 111 directions. The calculated peaks, overlooking the intersite interaction effects were located at $T_3 \sim T_1/2$ even though the experimental results were better characterized by a smaller value of $T_3$. Similar features have been found recently by Flint, Chandra and Coleman[33].

In the above discussion, the spin quantization axis is along the z direction, as is the external magnetic field. When the external field is perpendicular (spin quantization along z but the external field along x), we find a different characteristic response. The starting Hamiltonian is then H = $\Delta S_z^2 - \gamma S_x B$. The eigenvalues are $\Delta$, $1/2(\Delta \pm \sqrt{[\Delta^2+4(\gamma B)^2]}$ leading to:

$$\chi_1 = \frac{4\gamma^2}{\Delta} \left[\frac{1}{3\coth\left(\frac{1}{2\tau}\right)-1}\right] \qquad \text{Eqn.(5)}$$

and

$$\chi_3 = \frac{4\gamma^4}{\Delta^3} \frac{\coth\left(\frac{1}{2\tau}\right)\left[3\coth\left(\frac{1}{2\tau}\right)-(1+6\tau)\right]+2(\tau-1)}{\tau[3\coth\left(\frac{1}{2\tau}\right)-1]^2} \qquad \text{Eqn.(6)}$$

These susceptibilities are monotonic, they increase in magnitude with decreasing T, finally saturating at the lowest possible temperatures, on a scale determined by $\Delta$. The linear susceptibility $\chi_1$ remains positive and saturates at $2\gamma^2/\Delta$. The nonlinear susceptibility $\chi_3$ remains less than zero but saturates to $16\gamma^4/\Delta^3$. These functions are shown in the lower panel of fig.3 along with the experimental results on UPt$_3$.

It is useful to further compare this simple model with the overall temperature dependence of $\chi_3$ found experimentally. We do this in fig. 4 where we plot the values of $\chi_3$ normalized to their maximum values for three different systems measured by us, UPt$_3$, URu$_2$Si$_2$ and UPd$_2$Al$_3$. The temperature as represented along the x-axis likewise is scaled to the values of the single energy scale $\Delta$, different for each material obtained by using the experimental T$_1$ value in the theoretical result, T$_1$ = 0.69 $\Delta$. The individual values of $\Delta$ are provided in the figure. The comparison of the overall behavior of $\chi_3$ with the experimental results is extremely good but with notable discrepancies. The experimental values of $\chi_3$ (and $\chi_1$ as well) tend to fall outside of the theoretically calculated values in general. Since it is known that there are additional contributions to the susceptibilities not considered in our minimal model it is reasonable to expect this difference.

In conclusion, new experimental results on the strongly correlated metals, UPt$_3$, URu$_2$Si$_2$ and UPd$_2$Al$_3$ presented here taken together with the previous measurements of the nonlinear susceptibility in other heavy fermion systems indicate a universal scaling relation between the temperatures where maxima occur in the linear and the leading order nonlinear magnetic response to an applied field. Along with concurrent scaling of the metamagnetic critical field these correlations suggest a minimal model, consisting of two energy levels that approach each other under the influence of a magnetic field. The linear and nonlinear susceptibilities of this model have the maxima in their temperature dependence which (a) scale with the original energy separation and (b) follow T$_3$/ T$_1$=0.4. A peak in the next higher order susceptibility, $\chi_5$ can also be shown to follow from this model. These and other predictions will be discussed in a forthcoming extended paper. The surprising success of this simple model in accounting for the behavior of a seemingly large and diverse class of correlated metals lays the ground work for developing future microscopic theories.

Acknowledgements:  The authors wish to thank L. P. Gorkov and T.V. Ramakrishnan for many discussions and for generously sharing their vast insight into quantum magnetism.   Thanks are also due to Piers Coleman, Daniel Cox and Tony Leggett for useful advice and conversations. Work at the University of Virginia was made possible through grant NSF DMR 0073456. Research at UCSD was supported by the U.S. Department of Energy under Grant No. DE-FG02-04-ER46105.

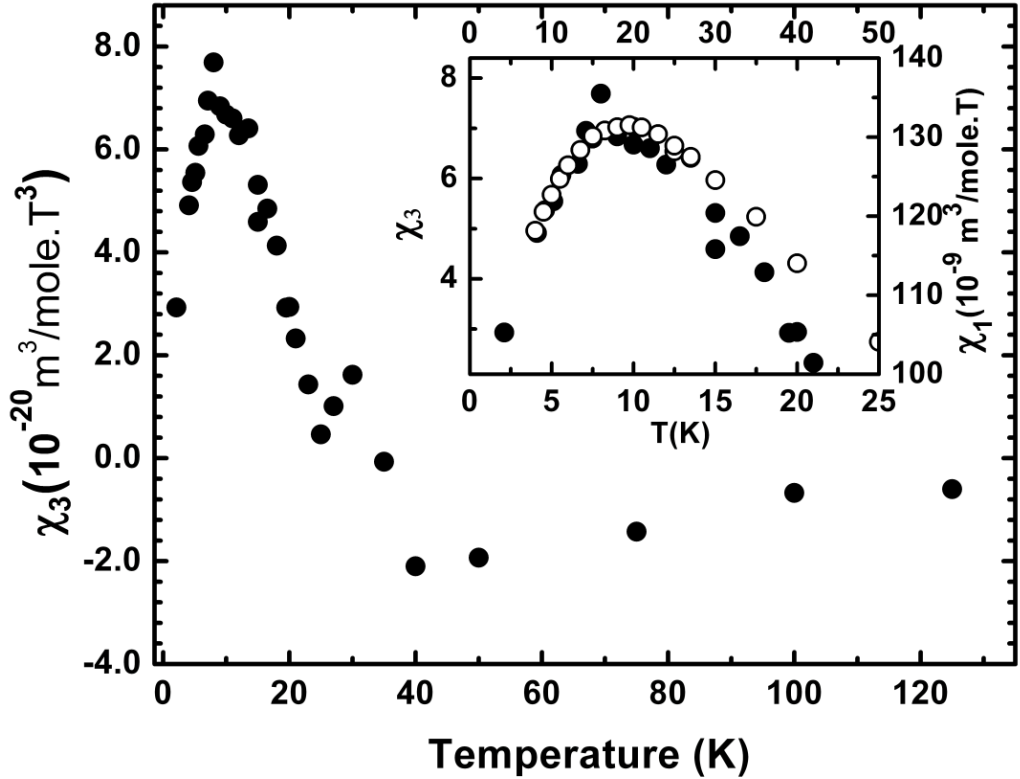

*Figure 1: The main part of the figure shows the non linear magnetic susceptibility in a single crystal of UPt₃ with field along the a-axis. The inset shows a comparison of both the linear and nonlinear susceptibilities. The bottom horizontal scale in the inset corresponds to $\chi_3$ and the top scale corresponding $\chi_1$ differs by a factor of two to illustrate the scaling of the two susceptibilities. The units for $\chi_3$ in the inset are the same as in the main figure.*

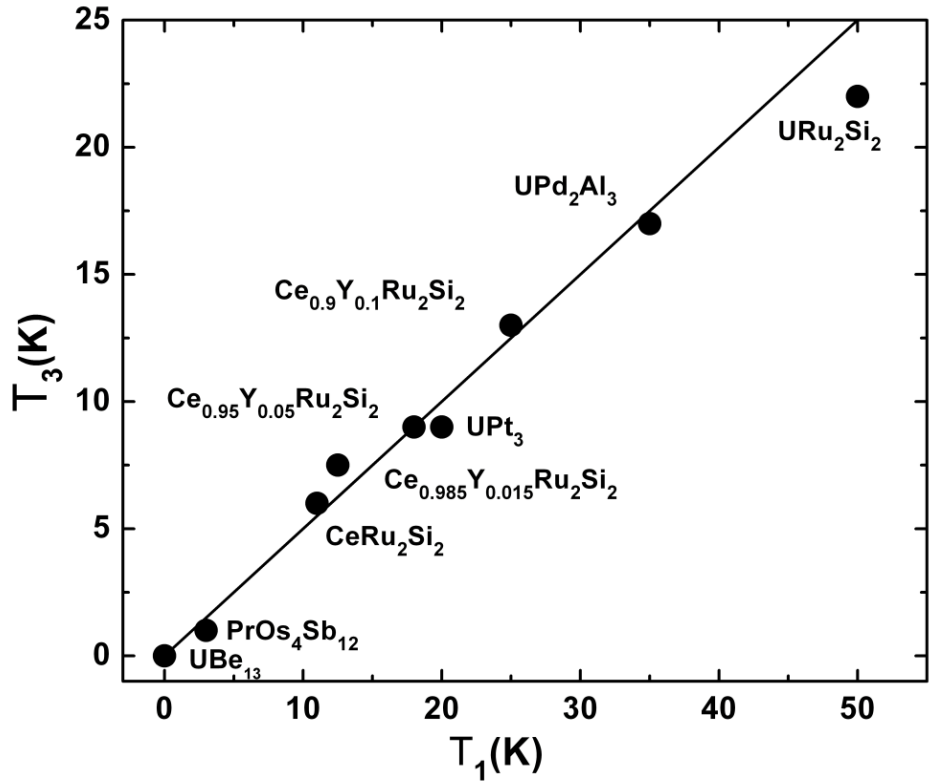

*Figure 2: Shows the scaling of the temperature of the maximum in the linear susceptibility, $\chi_1$ with the temperature of the maximum in $\chi_3$. We have used values for CeRu$_2$Si$_2$ (including alloys with Y) and PrOs$_4$Sb$_{12}$ as reported in refs. 20 and 23 respectively. The slope of the solid line shown is 0.5.*

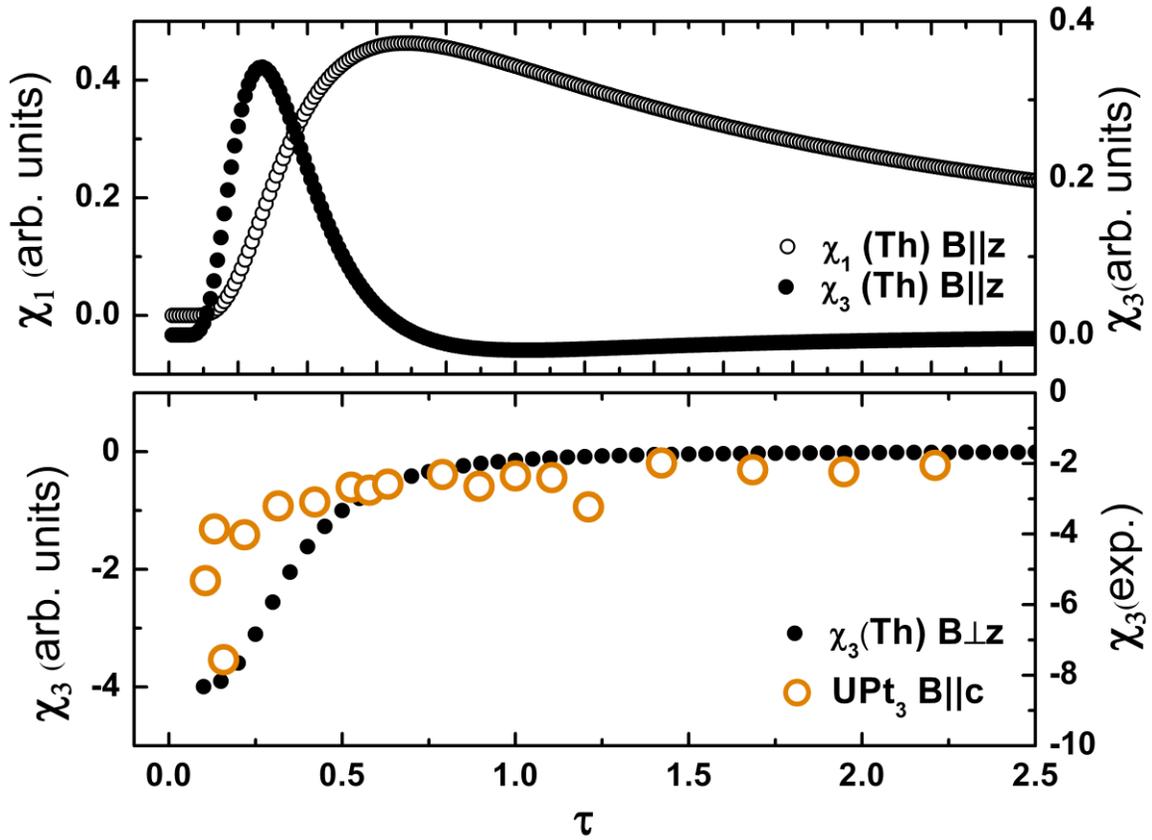

*Figure 3: The calculated linear and nonlinear magnetic susceptibility as a function of τ (temperature T normalized to the energy splitting Δ). These calculations do not include any exchange interaction. The upper panel is for the case when B is along the quantization axis. The lower panel is for the case B perpendicular to the quantization axis and shows only the nonlinear part. Experimental results are also shown for UPt₃. Here the units for χ₃ (lower panel - right hand scale) are the same as in fig.1.*

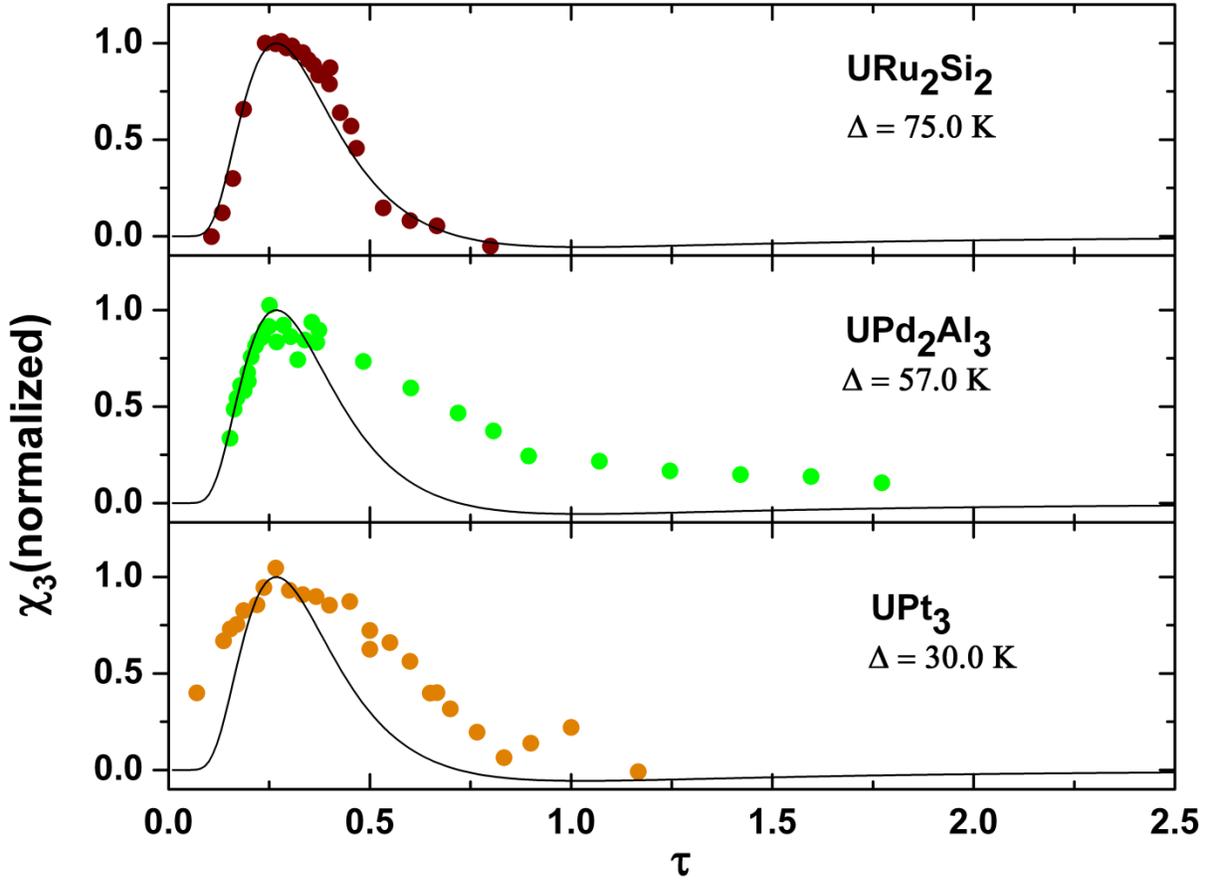

*Figure 4: Shows the experimental values of the nonlinear susceptibility, $\chi_3$, normalized to their maximum value plotted against the reduced temperature $\tau=T/\Delta$. The $\Delta$ values chosen for UPt$_3$, URu$_2$Si$_2$ and UPd$_2$Al$_3$ as mentioned above are obtained from the experimental values of $T_1$ the temperature where a maximum in the linear susceptibility occurs ( $\Delta=T_1/0.67$).*